# The Struggle with Academic Plagiarism: Approaches based on Semantic Similarity

Tedo Vrbanec[1], Ana Meštrović[2]
[1]Faculty of Teacher Education, University of Zagreb, Croatia
[2]Department of Informatics, University of Rijeka, Croatia
tedo.vrbanec@gmail.com, amestrovic@inf.uniri.hr

**Abstract** - Academic plagiarism is a serious problem nowadays. Due to the existence of inexhaustible sources of digital information, today it is easier to plagiarize more than ever before. The good thing is that plagiarism detection techniques have improved and are powerful enough to detect attempts of plagiarism in education. We are now witnessing efficient plagiarism detection software in action, such as Turnitin, iThenticate or SafeAssign. In the introduction we explore software that is used within the Croatian academic community for plagiarism detection in universities and/or in scientific journals. The question is - is this enough? Current software has proven to be successful, however the problem of identifying paraphrasing or obfuscation plagiarism remains unresolved. In this paper we present a report of how semantic similarity measures can be used in the plagiarism detection task.

**eywords** academic plagiarism, plagiarism detection, obfuscation plagiarism, natural language processing, semantic similarity

## I. INTRODUCTION

Academic plagiarism is nowadays one of the most pressing problems of the academic community. Many successful plagiarism detection tools and software products have been developed. However, the detection of paraphrasing or obfuscation plagiarism remains a challenge because most of the existing tools are only able to detect copy-paste cases of plagiarism. Plagiarism is not just the direct copying of one's text. It can be far more complicated if it is a case of paraphrasing or obfuscation. According to [1] high-obfuscation plagiarism can be realised by modifying original text by reduction, combination, paraphrasing, summarizing, restructuring, concept specification and concept generalization.

Due to this, there are serious drawbacks of systems such as TurnitIn or SafeAssign. More precisely, these tools cannot deal with the vocabulary problems such as synonymy, homonymy, hyperonymy and hyponymy. There are many various approaches in the area of natural language processing (NLP) that may offer a solution for this task.

However, there is still lot of room for improvement. This is why we are focused on paraphrasing and obfuscation plagiarism detection. We analyse various approaches that may identify paraphrasing and obfuscation by means of semantic similarity measures. Some are based on external knowledge resources such as WordNet or ontologies, and others are based on statistical NLP techniques.

The goal of this paper is to analyse the possibilities of existing semantic similarity measures and somehow classify existing approaches. This is just a preliminary study of this wide domain with the further goal of providing an extensive overview and classification of all semantic similarity measures of text and possible approaches in paraphrasing identification. In addition to this, we give a short review of the situation with academic plagiarism in Croatia. This is important because the struggle against academic plagiarism nowadays is on-going and there are frequent clashes.

In the first part of the paper we describe the problems of academic plagiarism. Furthermore, we give a brief review of academic anti-plagiarism efforts in Croatia. In the second part of the paper we are focused on various approaches for plagiarism detection based on semantic measures. After this we give an overview of other papers which try to resolve plagiarism detection problems with NLP techniques and semantic similarity. Finally, the sixth section contains a conclusion and possible directions for future research.

## II. ACADEMIC PLAGIARISM

Academic plagiarism in other words the plagiarism of digital text is most often the target of plagiarism during education and in academic papers. Academic plagiarism is a syntagma which indicates the plagiarism of a complete or part of documents of the following kinds: of programs in the source programming code, seminars, critical reviews, professional or scientific papers and non-literary books. The first tagmeme of syntagma - academically, points out that this kind of plagiarism most often appears in the academic community. At the same time it means that in the academic context plagiarism is a particularly worrying phenomenon which attracts the attention of all academic structures.

Below we provide a list of the methods of academic plagiarism and after that comment upon the anti-plagiarism efforts in the Croatian academic community.

*A.    ethods of Academic   lagiarism*

The most famous plagiarism software manufacturer TurnitIn [2], distinguishes academic plagiarism methods and research paper plagiarism methods. As academic plagiarism methods it lists:

- the submission of someone else's work as one's own, in order to fulfil a specified teaching obligation,
- the copying of words or ideas without giving credit to the original author,
- copying the majority of the words or ideas that compromises the work,
- submitting an already submitted work (e.g. from another colleague),
- not using quotation marks when quoting,
- giving incorrect data about sources,
- the use of someone else's sentences by using substitute words,
- using someone else's ideas without referencing.

TurnitIn also lists the research plagiarism methods [2, Pt. 1, p. 5]:

- "*Claiming authorship on a paper or research that is not one's own.*
- *Citing sources that were not actually referenced or used.*
-  *eusing previous research or papers without proper attri ution.*
- *Paraphrasing another's work and presenting it as one's own.*
-  *epeating data or te t from a similar study with similar methodology without proper attri ution.*
-  *u mitting a research paper to multiple pu lications.*
-  *ailing to cite or ac nowledge the colla orative nature of a paper or study*".

*B.   oftware solutions used in Academics*

In order that the academic community could effectively fight the modern plague in science and education – plagiarism, it is necessary [3, p. 123] (1) "to form warnings and measures in the education of students and scientists" … "at all levels of education" [4, p. 108] and (2) to develop or use existing software systems for plagiarism detection. According to [3], [4], the most widely used software products in the world are: iThenticate, SafeAssign and CrossCheck. All the papers checked by TurnitIn are stored in a database for further comparison. SafeAssign is optional in this regard. CrossCheck uses a large database of papers which have been handed over for the use of scholarly publishers affiliated to their CrossRef organisation. In return they can use the plagiarism detection system free of charge. Individuals do not generally have some cheap or free of charge choice. One of the possibilities is the Viper desktop application, which only works in the Windows environment.

It is known that with the software detection of plagiarism there still exists the insufficient detection of so-called intelligent plagiarism cases [3]: the plagiarism of ideas, complex paraphrasing, and plagiarism between multiple languages. This is trying to be resolved in existing commercial software systems with the adding of translation modules, however the solution is in a qualitative upturn: the detection of sematic similarities between documents.

*C. Academic Anti   lagiarism    fforts in Croatia*

In the EU and around the world, plagiarism is a very worrying phenomenon against which educational, preventative and restrictive methods are used. In 2015 The Council of Europe supported this effort by establishing the Pan-European Platform on Ethics, Transparency and Integrity in Education (ETINED), the aims of which priority activities are, amongst others, [5]: "Ethical behaviour of all actors in education" and "Academic integrity and plagiarism". In this context the European Commission led a project (2010-2013) in which Croatia did not participate: The Impact of Policies for Plagiarism in Higher Education Across Europe (IPPHEAE) "whose aim was to explore policies and systems of assuring academic integrity and deterring plagiarism in a system of higher education" [6, p. 5].

However, in Croatia as a member of the same EU, it is possible that the secretary of a parliamentary party, a rector of a university or a member of the Constitutional Court use plagiarism, yet in doing so are not sanctioned, what's more – they continue to perform their high office. And this is all despite the fact that in 2006 a Committee of Ethics in Science and Higher Education was established as a body of the highest legislative authority in Croatia (of Parliament).

TurnitIn has been used by the University of Rijeka since 2014 [3, p. 12] and immediately after by University of Osijek. The VERN Polytechnic and the Zagreb School of Management use TurnitIn, too [6, p. 12], as well as by individual faculties of the University of Zagreb, where there is unfortunately no joint financing, although the Rector announces it. In the meantime, individual faculties autonomously negotiate plagiarism checking services at their own expense. For example, the Faculty of Teacher Education in Zagreb used Ephorus for several years, however due to its high price and it crossed over to the cheaper solution: desktop only and Windows only application Plagiarism-Detector Personal.

The University of Split planned to begin using some software from 2015, but since the beginning of 2017 they have still not done so, due to the high prices. On the level of collaboration between Croatian universities, the president of the Rector's Assembly announced the collaboration of projects of mutual interest and a reduction of the burden of cost to each university.

Some unspecified plagiarism checking software is used by the School of Dental Medicine and the Faculty of Political Science [7, p. 2]. Furthermore, some scientific journals in Croatia use plagiarism detection software systems, e.g. Biochemia medica [8], the journal of the Croatian Society of Medical Biochemistry and Laboratory Medicine which both use iThenticate since 2013.

Members of professional committees for (re)accreditation from abroad are participating in the processes of the (re)accreditation of Croatian higher education institutions. Usage of plagiarism detection tools is one of the parameters for measuring the quality of professional and scientific work in academic institutions. This has prompted many of them to start using some plagiarism checking software or to consider options of using it. Surely the best solution for all of them is to unite, as many educational institutions as possible, in a common approach to suppliers, i.e. service providers, because in this way individual institutional usage is the cheapest, and further on, the database of documents – the corpus, as the basis for software comparison, becomes more complete. The current practice is that individual universities or faculties sign contracts with the software dealers or manufacturers, usually in three years' contracts, for a price that is of much greater than if they bought together and without less limitation of usage.

### III. SEMANTIC SIMILARITY

The concept of semantic similarity is fundamental and widely understood in many domains of natural language processing [9]. It can be defined as the degree of taxonomic proximity between terms [10]. The terms that can be also used instead are semantic proximity or semantic distance as an opposite concept. According to Resnik [11], semantic similarity represents a special case of semantic relatedness. He gives an example that cars and gasoline seem more closely related than cars and bicycles, but the latter pair is more similar.

Semantic similarity can be measured in terms of numerical score that quantifies similarity/proximity. In existing research's various semantic similarity measures (SSMs) have been defined and many semantic similarity computational models have been proposed. Semantic similarity measures have been widely used in many NLP and related fields such as text classification, information retrieval, information extraction, word sense disambiguation, machine translation, question answering, plagiarism detection, etc.

Semantic similarity can be calculated on different levels of granularity (between words, between paragraphs or between whole texts/documents).

In the case of plagiarism detection, semantic similarity should be expressed on the text level as the final result. The score of semantic similarity between suspected document and one or more other documents may indicate the existence of plagiarism. The whole procedure of plagiarism identification, semantic similarity can be calculated on the sentence and paragraph level.

### IV. SEMANTIC SIMILARITY MEASURES

In this section we will give an overview of different measures and approaches in measuring the semantic similarity of texts. Semantic similarity measures are defined in the domain of NLP for various tasks. One of the first application of SSMs was obtained in 1970s seventies for the task of information retrieval [12].

There are various approaches in detecting semantic similarity. According to [13], there are two main approaches in measuring the semantic similarity of texts: corpora-based and knowledge-based. There is also a class of ontology-based semantic similarity measures extensively described in [10]. These measures also belong to the class of knowledge-based measures. Here we present another approach in which we classify measures it two categories: SSMs on the concept/word level and SSMs on the document/text level (Fig. 1).

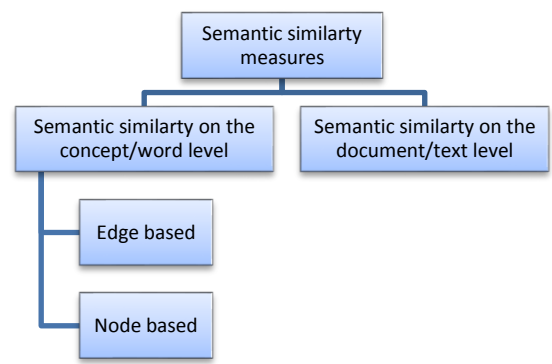

igure .  imple classification of semantic similarity approaches

#### A.  emantic similarity on the concept word level

Semantic similarity on the concept or word level is based on the hierarchy between concepts or words. It is usually defined for the taxonomy such as WordNet or for some more extensive ontology [10]. These measures assume as input a pair of concepts or words, and return a numeric value indicating their semantic similarity. Based on word similarities it is possible to compute the similarities of texts according to variously defined equations [13].

If we have an ontology or "IS-A" taxonomy with a set of concepts $C$. The similarity between two concepts can be computed by taking into account the edges (edge-based measures) or by taking into account the nodes (node-based measures).

Edge-based measures estimate the similarity of two concepts according to how near these two concepts are. The simplest way to measure semantic similarity of two concepts $c_1$ and $c_2$ is to estimate the distance of two concepts as the shortest-path between them [14]. There are more sophisticated approaches that take into account the possible weights of edges between two concepts. In most cases, the semantic similarity of two concepts is estimated as a function of the depth of the Least Common Ancestor (LCA) or Least Common Subsumer (LCS) [10]. An example of this approach is Wu and Palmer metrics [15] with a very simple similarity measure of two concepts:

$$sim_{WP}(c_1, c_2) = \frac{2depth(LCA_{c_1,c_2})}{depth(c_1) \cdot dept(c_2)}.$$

Node based approaches can be divided into two categories: a feature-based and one based on information theory.

In a feature based approach a concept is described with a set of features. Thus, two concepts can be compared in terms of classical binary or distance measures. An example of this approach is the Concept-Match similarity measure defined by Maedche and Staab [16]:

$$sim_{CMatch}(c_1, c_2) = \frac{|F(c_1) \cap F(c_2)|}{|F(c_1) \cup F(c_2)|}.$$

Approaches based on information theory are based on Shannon's theory. The similarity of concepts is defined according to the amount of information they share. Resnik define a measure:

$$sim_{Resnik}(c_1, c_2) = \max_{c \in S(c_1,c_2)}[-\log p(c)],$$

where $S(c_1, c_2)$ is the set of concepts that subsume both concepts $c_1$ and $c_2$.

Pointwise Mutual Information is proposed by Turney [17]. It is based on word co-occurrence using counts collected over very large corpora. For two words $w_1$ and $w_2$, their PMI-IR is measured as:

$$sim_{PMI}(w_1, w_2) = log_2 \frac{p(w_1, w_2)}{p(w_1) \cdot p(w_2)}.$$

There are many other concept level measures, named by their authors, amongst whom the important ones are: Leacock & Chodorow, Lesk, Wu and Palmer, Resnik, Lin, Jiang and Conrath, Zhong, Nguyen and Al-Mubaid, Caviedes and Cimino, etc.

*B. emantic similarity on the document te t level*

Similarity measures on the document or text level are mainly based on the approaches developed in the NLP domain. Most of these approaches have their roots in machine learning. In this section we briefly describe the most used approaches for calculating the semantic similarity between two documents.

One of the first approaches in measuring semantic similarity between documents was vector space model (VSM) originally proposed by Salton et al. [18] for the task of information retrieval domain.

In the VSM each document is a point in an $n$-dimensional space. For a given set of documents $D = \{D_1, D_2, ..., D_k\}$, a document $D_i$ can be represented as a vector $D_i = (w_{i1}, w_{i2}, ..., w_{in})$. In the classical word-based VSM, each dimension corresponds to one term or word from the document set. Weights may be determined by using various weighting schemes; TF-IDF is usually used in the word-based VSM. The similarity between two documents $D_i$ and $D_j$ can be calculated as cosine similarity:

$$sim_{cos} = \frac{D_i \cdot D_j}{\|D_i\| \cdot \|D_j\|}.$$

The main drawbacks of this model are high dimensionality, sparseness and vocabulary problems. Therefore, there are various modifications and generalisations of this classical version of the VSM.

Another approach is Latent semantic analysis (LSA) proposed by Landauer (1998). It also exploits the vector space model, but in this approach it uses a dimension reduction technique known as Singular Value Decomposition (SVD) of the initial matrix. In this way LSA overcomes the high dimensionality and sparseness of the standard SVM model. The similarity between two documents is again calculated as a cosine similarity as in equation or some other similarity measure.

One more recent approach is deep learning. Similarly with the previous methods, in deep learning documents or texts can be represented as vectors by the using document to vector technique (doc2vec). Moreover, words are also represented as vectors by using the word to vector strategy (word2vec). There are variations in learning word vector representation; one is the matrix decomposition method such as LSA; another is context-base methods such as skip-grams, a continuous bag of words. Furthermore, there is an unsupervised algorithm that learns representation for documents or smaller samples of texts (paragraphs, sentences). At the end, vectors can be compared using cosine similarity or some other similarity measure.

There are also other similar measures for text similarity such as Explicit Semantic Analysis (ESA), Salient Semantic Analysis (SSA), Distributional Similarity, Hyperspace Analogues to Language, etc.

V. RELATED WORK

There are attempts to identify plagiarism by semantic similarity measures. Researchers have approached the problem of determining plagiarism through semantic similarity in a multitude of ways. Many have used ontologies, usually WordNet with some additions like fuzzy similarity measures, a combination of WordNet with morphological and syntactic analysis, machine learning from examples or with hashing, etc. Others have turned to intrinsic methods, citation-based plagiarism detection, graphical representation, natural language processing, deep learning and even multidimensional approach.

In [19] Tsatsaronis et al. (2010) presented a semantic-based approach to text-plagiarism. Their approach is based on WordNet and the Wikipedia as knowledge bases which can resolve the problems of synonymy and hyponymy/hypernymy. Similarly, in [20] Fernando and Stevenson (2008) used WordNet as the knowledge base in the proposed algorithm for paraphrase identification.

Shenoy et al. (2012) [21, p. 59] "proposes an automatic system for semantic plagiarism detection based on ontology mapping". They invented an algorithm which is capable of learning ontology from documents using Web Ontology Language OWL and then applying it to ontology mapping to "detect correspondences between the various entities" [21, p. 60].

WordNet has been used by more than a few researchers. Al-Shamery and Gheni (2016) [22] consider that finding synonyms (over WordNet) on the same place in comparing documents is to be the proof of semantic plagiarism.

Alzahrani and Salim (2010) [23] presents "plagiarism detection method using a fuzzy semantic-based string similarity approach". They pull potential source documents using shingling and Jaccard coefficient, then compare them to the sentence granularity, simultaneously computing fuzzy degree of similarity with the help of WordNet a (different words gets 0, WordNet synonyms gets 0.5, the same word gets 1 fuzzy degrees).

Marsi and Krahmer (2010) [24] suggested the semantic similarity method for analysing comparable text that relies on a combination of morphological and syntactic analysis, WordNets, and machine learning from examples. They analyse semantic similarity between sentences "by aligning their syntax trees, where each node is matched to the most similar node in the other tree (if any)" [24].

Czerski et al. (2015) [25] proposed using an algorithm based on sentence hashing but the approach was combined with replacing some word for some representative ones using synonyms and the WordNet, thesaurus and "IS A" ontology, so the number of sentences in the document is reduced and the remaining sentences consist only from lemmas.

Eissen and Stein (2006) [26, pp. 566–567] used the intrinsic method of semantic similarity discovering: stylometry analysis, using five categories of stylometric features: "(i) text statistics, which operate at the character level, (ii) syntactic features, which measure writing style at the sentence-level, (iii) part-of-speech features to quantify the use of word classes, (iv) closed-class word sets to count special words, and (v) structural features, which reflect text organization." They also introduced a new stylometric measure: averaged word frequency class, as "the most powerful concept with respect to intrinsic plagiarism detection" [26, p. 567].

Gipp et al. (2013) [27, p. 1119] noticed that plagiarists "usually do not substitute or rearrange the citations copied from the source document", so they developed several Citation-based Plagiarism Detection algorithms using citation patterns within scientific documents "as a unique, language-independent fingerprint to identify semantic similarity" with reasonable success in detecting disguised plagiarism.

Osman et al. (2010) designed a method of detecting plagiarism based on graph representation. For two documents their method [28, p. 36] "build the graph by grouping each sentence terms in one node, the resulted nodes are connected to each other based on order of sentence within the document, all nodes in graph are also connected to top level node" which is "formed by extracting the concepts of each sentence terms and grouping them in such node".

Chong et al. (2010) [29] applied several NLP techniques on short paragraphs to analyse the structure of the text to automatically identify plagiarised texts. They proved that NLP techniques can improve the accuracy of detection tasks, although there remain challenges such as multilingual detection, synonymy generalisation (word sense disambiguation) and sentence structure generalisation.

Gharavi et al. (2016) [30, p. 1] proposed a "deep learning based method to detect plagiarism" in the Persian language. "In the proposed method, words are represented as multi-dimensional vectors, and simple aggregation methods are used to combine the word vectors for sentence representation. By comparing representations of source and suspicious sentences, pair sentences with the highest similarity are considered as the candidates for plagiarism. The decision on being plagiarism is performed using a two level evaluation method" [30, p. 1].

Mihalcea et al. (2006) presented a method that outperforms a vector-based similarity approach for measuring the semantic similarity of texts using two corpus-based and six knowledge-based measures of word similarity which they used "to derive a text-to-text similarity metric" [13, p. 775].

In [1] Kong et al. (2014) tried to detect high obfuscation plagiarism with a Logical Regression model. The proposed model integrated lexicon features, syntax features, semantics features and structure features which are extracted from suspicious documents and source documents.

VI. DISCUSSION AND CONCLUSION

In this paper we want to describe our preliminary research on semantic similarity measures and their possible usage for paraphrasing detection in the task of plagiarism identification. We analyse some existing measures for quantifying the semantic similarity of texts. There is a plethora of measures and approaches proposed and we divided them into two basic categories with some subcategories. All these measures are defined for different purposes in NLP domain. However, according to the related work, it is obvious that some of these measures can be utilized in the task of plagiarism detection.

One possible approach is based on external knowledge represented in some formalism. External knowledge can be represented in ontology or simpler taxonomies such as WordNet. However, the formalism is not limited to these classical ontologies/taxonomies; it can be any kind of graph representation of lexical relations [31]. Another approach is to use statistical methods designed in the domain of NLP. In the section about related work we present all the approaches that have been used recently for plagiarism detection.

There may be one drawback of the approach based solely on the semantic similarity measures. We need to point out that, perhaps these semantic measures are not enough, and that they can be combined with classical approaches that may identify copy-paste plagiarism.

For further research we plan to experiment with the NOK method [32] or some other graph based formalism for lexical relation representation [33]. Additionally, we would like to experiment with the approach that we try with an ontology-based information retrieval in which the classical VSM is projected onto a smaller vector space [34].

VII. REFERENCES

[1] L. Kong, Z. Lu, H. Qi, and Z. Han, "Detecting High Obfuscation Plagiarism: Exploring Multi-Features Fusion via Machine Learning," *nt. . u and e ervice ci. Technol.*, vol. 7, no. 4, pp. 385–396, 2014.

[2] Turnitin Europe, *lagiarism in a Digital orld series*. Turnitin, 2016.

[3] S. Lampret, V. Pupovac, and M. Petrovečki, "Računalni programi i programske usluge za otkrivanje plagiranja u znanosti i obrazovanju," *D* , vol. 18, no. 98/99, 2012.

[4] K. Baždarić, V. Pupovac, L. Bilić-Zulle, and M. Petrovečki, "Plagiarism as a violation of scientific and academic integrity," 2009.

[5] Council of Europe, "ETINED - The Pan-European Platform on Ethics, Transparency and Integrity in Education," 2017. [Online]. Available: http://www.coe.int/en/web/ethics-transparency-integrity-in-education. [Accessed: 22-Jan-2017].

[6] T. Birkić, D. Celjak, M. Cundeković, and S. Rako, "Izvještaj: Analiza softvera za otkrivanje plagiranja u znanosti i obrazovanju," Zagreb, Sep. 2016.

[7] S. Vuković and B. Kopić, "Plagiranje - Sveučilište u Zarebu još nije uvelo sustav provjere radova," *lo al* , Zagreb, p. 24, Nov-2016.

[8] V. Supak Smolcic and A.-M. Simundic, "Biochemia Medica has started using the CrossCheck plagiarism detection software powered by iThenticate," *Biochem. edica*, pp. 139–140, 2013.

[9] J. O. Shea, Z. Bandar, K. Crockett, and D. Mclean, "A Comparative Study of Two Short Text Semantic Similarity Measures," *Artif. ntell.*, vol. 4953, pp. 172–181, 2008.

[10] S. Harispe, D. Sánchez, S. Ranwez, S. Janaqi, and J. Montmain, "A framework for unifying ontology-based semantic similarity measures: A study in the biomedical domain," *. Biomed. nform.*, vol. 48, pp. 38–53, Apr. 2014.

[11] P. Resnik, "Semantic Similarity in a Taxonomy: An Information-Based Measure and its Application to Problems of Ambiguity in Natural Language," *. Artif. ntell. es.*, vol. 11, pp. 95–130, May 2011.

[12] G. Salton and M. E. Lesk, "Computer Evaluation of Indexing and Text Processing," *. AC* , vol. 15, no. 1, pp. 8–36, Jan. 1968.

[13] R. Mihalcea, C. Corley, and C. Strapparava, "Corpus-based and knowledge-based measures of text semantic similarity," 2006, vol. 6, pp. 775–780.

[14] R. Rada, H. Mili, E. Bicknell, and M. Blettner, "Development and Application of a Metric on Semantic Nets," *Trans. yst. an Cy ern.*, vol. 19, no. 1, pp. 17–30, 1989.

[15] Z. Wu and M. Palmer, "Verbs semantics and lexical selection," in *roceedings of the nd annual meeting on Association for Computational inguistics* , 1994, pp. 133–138.

[16] a Maedche and S. Staab, "Comparing ontologies-similarity measures and a comparison study," *roc. of A* , no. 408, 2002.

[17] P. Turney, "Mining the Web for Synonyms: PMI-IR Versus LSA on TOEFL," 2001.

[18] G. Salton, A. Wong, and C.-S. Yang, "A vector space model for automatic indexing," *Commun. AC* , vol. 18, no. 11, pp. 613–620, 1975.

[19] G. Tsatsaronis, I. Varlamis, A. Giannakoulopoulos, and N. Kanellopoulos, "Identifying free text plagiarism based on semantic similarity," in *roceedings of the th nternational lagiarism Conference*, 2010, no. i.

[20] S. Fernando and M. Stevenson, "A semantic similarity approach to paraphrase detection," in *roceedings of the th Annual esearch Collo uium of the U pecial nterest roup for Computational inguistics*, 2008, pp. 45–52.

[21] M. K. Shenoy, K. C. Shet, and U. D. Acharya, "Semantic Plagiarism Detection System Using Ontology Mapping," *Adv. Comput. An nt. .*, vol. 3, no. 3, pp. 59–62, May 2012.

[22] E. S. Al-Shamery and H. Q. Gheni, "Plagiarism Detection using Semantic Analysis," *ndian . ci. Technol.*, vol. 9, no. 1, Feb. 2016.

[23] S. Alzahrani and N. Salim, "Fuzzy semantic-based string similarity for extrinsic plagiarism detection," *Braschler and arman*, 2010.

[24] E. Marsi and E. Krahmer, "Automatic analysis of semantic similarity in comparable text through syntactic tree matching," 2010, pp. 752–760.

[25] D. Czerski, P. Lozinski, A. Cacko, R. Szmit, and B. Tartanus, "Fast plagiarism detection in large scale data," 2015.

[26] S. M. Zu Eissen and B. Stein, "Intrinsic plagiarism detection," Springer, 2006, pp. 565–569.

[27] B. Gipp, N. Meuschke, C. Breitinger, M. Lipinski, and A. Nürnberger, "Demonstration of citation pattern analysis for plagiarism detection," 2013, pp. 1119–1120.

[28] A. H. Osman, N. Salim, and M. S. Binwahlan, "Plagiarism detection using graph-based representation," *. Comput.*, vol. 2, no. 4, 2010.

[29] M. Chong, L. Speciali, and R. Mitkov, "Using natural language processing for automatic detection of plagiarism," 2010.

[30] E. Gharavi, K. Bijari, H. Veisi, and K. Zahirnia, "A Deep Learning Approach to Persian Plagiarism Detection," 2016.

[31] M. Pavlić, A. Meštrović, and A. Jakupović, "Graph-based formalisms for knowledge representation," 2013, vol. 2, pp. 200–204.

[32] A. Jakupović *et al.*, "Comparison of the Nodes of Knowledge method with other graphical methods for knowledge representation," 2013, pp. 1004–1008.

[33] A. Meštrović and M. Čubrilo, "Monolingual Dictionary Semantic Capturing Using Concept Lattice," *nt. ev. Comput. oftw.*, vol. 6, no. 2, pp. 173–184, 2011.

[34] A. Meštrović and A. Calì, "An Ontology-Based Approach to Information Retrieval," in *emanitc eyword ased earch on tructured Data ources*, 2017, pp. 150–156.